\documentclass[preprint,showpacs,preprintnumbers,pra]{revtex4}
\usepackage{graphics}
\usepackage{longtable}
\usepackage{graphicx}
\usepackage{dcolumn}
\usepackage{bm}
\usepackage{amsmath}
\usepackage{amsfonts}
\usepackage{amssymb}

\begin{document}

\title{xOEP calculations using Slater-type basis functions: 
atoms and diatomic molecules}
\author{J. J. Fern\'{a}ndez}
\email{jjfernandez@fisfun.uned.es}
\affiliation{Departamento de F\'{\i}sica Fundamental, UNED, 
Apartado 60.141, E-28080 Madrid (Spain). }
\author{J. E. Alvarellos}
\affiliation{Departamento de F\'{\i}sica Fundamental, UNED, 
Apartado 60.141, E-28080 Madrid (Spain). }
\author{P. Garc\'{i}a-Gonz\'{a}lez}
\affiliation{
Departamento de F\'{\i}sica Te\'{o}rica de la Materia Condensada, 
Universidad Aut\'{o}noma de Madrid. Cantoblanco. 
E-28049 Madrid (Spain).}
\author{M. Filatov}
\affiliation{Department of Theoretical Chemistry, Zernike Institute for Advanced
Materials, 
University of Groningen, Nijenborgh 4, 9747AG Groningen (Netherlands).}

\date{\today }

\begin{abstract}
\noindent
The exchange-only optimized effective potential method is 
implemented with the use of Slater-type basis functions, 
seeking for an alternative to the standard methods of 
solution with some computational advantages.
This procedure has been tested in a small group 
of closed shell atoms and diatomic molecules, for which 
numerical solutions are available. 
The results obtained with this implementation have been 
compared to the exact numerical solutions and to the results 
obtained when the optimized effective equations are solved 
using the Gaussian-type basis sets. 
This Slater-type basis approach leads to a more compact 
expansion space for representing the potential of the 
optimized effective method 
and to considerable computational savings when compared to 
both the numerical solution and the more traditional one 
in terms of the Gaussian basis sets.
\end{abstract}

\maketitle

\section{Introduction}%
\label{Sec:Intro}

In practical implementations of the Kohn-Sham density 
functional theory \cite{1964HK,1965KS}
the exchange-correlation (XC) energy is usually described 
by a suitable explicit functional of the electron density 
and the parameters characterizing the density inhomogeneity.  
A very promising step to improve the accuracy of density 
functional calculations relies on the use of orbital-dependent 
density functionals, in which the XC energy is expressed 
as an explicit functional of Kohn-Sham (KS) orbitals
\cite{2008KK}. 
A rigorous approach to implement the orbital-dependent 
functionals within the KS formalism is the optimized 
effective potential (OEP) method %
\cite{2008KK, 1953SH, 1976TS, 1999ED}, 
where the XC potential is described by a local 
multiplicative term and the total energy functional 
is orbital-dependent.
By virtue of the Hohenberg-Kohn theorems \cite{1964HK}, 
the OEP solution is equivalent to the minimization 
of the total electronic energy with respect to the 
density \cite{1982SGP}. 
In the case of the exchange-only 
--- i.~e., the Hartree-Fock (HF) ---
energy functional, the corresponding OEP method (xOEP) was first 
formulated by Sharp and Horton \cite{1953SH} and numerically 
solved in real space for atoms by Talman and Shadwick \cite{1976TS}. 
Applications of real space xOEP formalism to atoms, molecules 
and solids, thanks to the use of pseudopotentials, have been 
reported in the literature %
\cite{1993EV,2009MKK,1999SMMVG,2008KKM,2009ES,2009MAEKK,%
2009Engel}.
A number of approximations to the xOEP method, 
such as the method of Krieger-Li-Iafrate %
\cite{1992KLI, 1993LKI},  
the local HF \cite{2001DSG} 
and the effective local potential \cite{2006SSD} methods 
have been developed.

The real space resolution of the xOEP method has only been  
efficiently applied to highly-symmetric systems, 
such as spherically symmetric atoms and diatomic molecules 
\cite{1976TS,2009MKK}. 
Application of the xOEP formalism to polyatomic molecules requires 
its formulation in terms of basis sets suitable for molecular 
calculations. 
Currently, there exist several formulations of the xOEP method in 
terms of basis sets of local Gaussian-type (GTO) functions
\cite{1999IHB,2007KF,2008KF,2007HGDSG,1999G}.
The most popular implementation of the xOEP formalism employs two 
different basis sets, 
one for the expansion of the KS orbitals and 
another one for representing the local multiplicative 
potential \cite{2007HGDSG,2002YW, 2008GHJL}. 
Within this approach, a special care must be taken when selecting 
the auxiliary basis set for the potential, thus leading 
to the concept of a balanced basis set firmly connected to the 
orbital basis set \cite{2008GHJL}. 
Alternatively, a set of the products of the occupied and virtual 
KS orbitals can be employed for the solution of the 
xOEP equations for the local potential %
\cite{2007KF, 1999IHB, 2001CN}. 

The computational complexity of the xOEP method in a basis set 
representation depends critically on the size of the orbital 
expansion basis set. 
For obtaining faithful solutions of the xOEP equations, the 
orbital basis set should support the linear dependence in 
the space of the occupied-virtual orbital products %
\cite{1999G, 1999IHB, 1983H}.
With the use of GTO basis functions, this requirement leads to 
very large orbital basis sets with 
hundreds of basis functions even for small molecules. 
On the other hand, the use of Slater-type (STO) basis functions 
give considerably more compact orbital basis sets which 
can be beneficial for the application of the xOEP formalism, 
and a very efficient implementation of quantum chemical formalisms 
with Slater-type basis functions has been achieved in the 
SMILES suite of programs \cite{2008SMILES}. 

It is the primary purpose of the present work to implement the 
xOEP method within the SMILES package and to analyze the advantages 
which can be obtained from the use of the STO basis sets, 
testing it for a small number of atoms and diatomic molecules, 
for which both numerical solutions \cite{1993EV, 2009MKK} 
and STOs are available \cite{2008SMILES}.
In this work we employ the xOEP algorithm outlined in 
Refs.~[\onlinecite{2007KF}] and [\onlinecite{2008KF}] 
and the xOEP equations, formulated in terms of the STOs.
The solution of the xOEP equation will be carried out 
through the truncated singular value decomposition (TSVD) 
technique. 
It will be demonstrated that the use of the Slater-type basis 
sets leads to considerable computational savings in every 
step of the selfconsistent procedure, without deteriorating 
the accuracy of the calculated xOEP total and orbital 
energies.

\section{Theory}
\label{Sec:Theory}

In this section the main features of the OEP method will be 
outlined. 
In the OEP method one seeks for a local multiplicative 
potential $V_{\sigma}({\mathbf r})$ such that its 
eigenfunctions (atomic units will be used in this paper) 
\begin{equation} \label{eigen}
\left( -\frac{1}{2} \nabla^{2} + V_{\sigma}({\mathbf r}) \right) 
\phi_{p\sigma} = 
\epsilon _{p\sigma} \phi_{p\sigma}
\end{equation}
minimize the total energy functional given by 
\begin{eqnarray} \label{Etot}
\nonumber
 E^{OEP}\left[ \{ \phi_{i\sigma} \} \right] 
 & = & 
\sum_{\sigma} \sum_{i} \int \phi_{i\sigma}^{*} 
( {\mathbf r}) \left( -\frac{1}{2} \nabla^{2} \right) 
\phi_{i\sigma}( {\mathbf r}) d{\mathbf r}  \\
& + & 
\int \rho( {\mathbf r}) V_{ext} ( {\mathbf r})  d{\mathbf r} 
+  
\frac{1}{2} \int \rho( {\mathbf r}) 
\int \rho( {\mathbf r}') \frac{1}{|{\mathbf r}-{\mathbf r}'|} 
d{\mathbf r} d{\mathbf r}' 
+
E_{xc}\left[ \{ \phi_{i\sigma} \} \right], 
\end{eqnarray}
where 
$\sum_{i} $ runs over occupied orbitals 
and $\sum_{\sigma}$ over the spin, 
being  
$\rho( {\mathbf r}) = \sum_{i} \sum_{\sigma} |\phi_{i\sigma}|^{2}$ 
the electron density and 
$E_{xc}\left[ \{ \phi_{i\sigma} \} \right]$ 
the orbital-dependent exchange-correlation energy. 
Here and below we use indices $i,j, ...$ for occupied 
orbitals, $a,b,...$ for unoccupied orbitals and 
$p,q,...$ for general (i.e. occupied or unoccupied) 
orbitals. 
In the case of the exchange-only formalism, xOEP, the 
XC energy 
$E_{xc}\left[ \{ \phi_{i\sigma} \} \right]$ 
is replaced with the HF exchange energy,
\begin{equation} \label{xenergy}
E_{x}\left[ \{ \phi_{i\sigma} \} \right] = 
- \frac{1}{2} \sum_{i,j} \sum_{\sigma} 
\int \frac{  \phi_{i\sigma}^{*}( {\mathbf r}) 
\phi_{j\sigma}( {\mathbf r})  \phi_{j\sigma}^{*}( {\mathbf r}')  
\phi_{i\sigma}( {\mathbf r}')}{|{\mathbf r}-{\mathbf r}|'}.
\end{equation}
The local multiplicative potential
\begin{equation}
V_{\sigma}({\bf r}) = 
V_{ext}({\bf r}) + 
\int \frac{\rho( {\mathbf r'})}{|{\mathbf r}-{\mathbf r'}|} d{\mathbf r'}
+ V_{\sigma}^{x}({\bf r})
\end{equation}
\noindent
is splitted into the external potential 
$V_{ext}({\bf r})$ (i.~e. the potential due to the nuclei), 
the Coulomb potential of the electron cloud and the local exchange 
potential $V_{\sigma}^{x}({\bf r})$. 

The xOEP equations in a basis set representation are 
obtained from the minimization of the total 
energy presented in Eq.~(\ref{Etot}) with respect to the local 
potential $ V_{\sigma}({\bf r})$ 
\cite{2008GHJL,2007HBY}, 
being this minimization equivalent to the minimization 
respect to the density, by virtue of the Sham-Schl\"{u}ter 
condition \cite{1982SGP} and the Hohenberg-Kohn 
theorems \cite{1964HK}. 

An expansion of the exchange part of the local potential 
is assumed in terms of an appropiate set of functions 
\cite{2002IHB, 2002YW, 2007HGDSG, 2007KF, 2008GHJL, 2007HBY}, 
\begin{equation} \label{vxspan}
V^{x}_{\sigma}({\mathbf r}) = 
\sum_{\mu} \tilde{w}_{\mu \sigma} f_{\mu}({\bf r}) ,
\end{equation} 
where $\sigma$ labels the spin, 
$\mu$ labels all needed values of the basis index 
and $\tilde{w}_{\mu \sigma}$ are the expansion coefficients 
of the exchange potential in this basis.
Following the literature \cite{2007KF} the expansion functions are 
conveniently defined as 

\begin{equation}  \label{gmu}
f_{\mu \sigma}({\bf r}) = 
\int \frac{g_{\mu \sigma}({\bf r}')}{|{\bf r}-{\bf r}'|} d{\bf r}'
\end{equation}
where $g_{\mu \sigma}({\bf r})$ are square integrable functions. 
Note that this definition implies that the expansion functions 
$f_{\mu \sigma}({\bf r})$ 
are not necessarily square integrable; 
this is not a problem as the local potential does not 
satisfy this condition. 

The requirement that the total energy to be stationary under 
the variations of the local potential, 
i.~e. $ \delta E^{xOEP} / \delta V_{\sigma}({\mathbf r}) = 0 $, 
is then equivalent to finding a minimum of the total xOEP 
energy with respect to the set of the expansion coefficients 
of the the local potential, 
$\{ \tilde{w}_{\mu \sigma} \}$. 
If we work with a real orbital basis 
and if we introduce a scalar product of two 
functions $h$ and $l$ of our expansion space as 
\begin{equation}
(h|l) = \int \int h({\mathbf r}) \frac{1} {|{\bf r}-{\bf r}'|} %
l({\mathbf r}') d{\mathbf r} d {\mathbf r}' ,
\end{equation}
the minimization of the xOEP energy 
--- i.~e. Eq.~(\ref{Etot}) with $E_{xc}$ defined as the 
Fock exchange energy given in Eq.~(\ref{xenergy}) --- 
leads to the equation \cite{2008KF}
\begin{equation}
\label{eqPot}
\frac{\partial E^{xOEP}}{\partial \tilde{w}_{\mu \sigma}} = 
2 \sum_{ia} \frac{\left(g_{\mu \sigma}|\phi_{a\sigma}\phi_{i\sigma}\right)}
{\varepsilon_{a \sigma} - \varepsilon_{i \sigma}}
\int \int d{\mathbf r} d{\mathbf r}' \phi_{i\sigma}({\mathbf r}') 
\left[ V^{x}_{\sigma}({\mathbf r}') - 
V_{\sigma}^{x,nl} ({\mathbf r}, {\mathbf r}') \right]  
\phi_{a\sigma} ({\mathbf r}')  
 = 0 .
\end{equation}
Here the non-local potential,
$ V_{\sigma}^{x,nl} ({\mathbf r}, {\mathbf r}') $, 
is defined as 
\begin{equation}
V_{\sigma}^{x,nl} ({\mathbf r}, {\mathbf r}') 
\; \phi_{j \sigma}({\mathbf r}')
=
\frac{\delta E_{x} \left[ \{ \phi_{i \sigma} (\mathbf{r}) \} \right] }%
{\delta \phi_{j \sigma}(\mathbf{r}') } ,
\end{equation}
being $\{ \phi_{q \sigma} \}$ the solutions of 
Eq.~(\ref{eigen}). 

Using the matrix
\begin{equation}
M^{\sigma}_{\mu,jb} = 
\frac{
\left( g_{\mu \sigma} | \phi_{b\sigma} \phi_{j\sigma} \right) } %
{ \sqrt{ \varepsilon_{b \sigma}- \varepsilon_{j \sigma} } }
\end{equation}
in Eq.~(\ref{eqPot}) we get a matrix equation equivalent 
to the minimization of the xOEP energy, 
\begin{equation}
\label{nablaOEP}
\nabla_{ \mathbf{\tilde{w}} } E^{xOEP} = %
2 \mathbf{MM}^{\dagger} \mathbf{\tilde{w}} - %
2 \mathbf{M} \mathbf{w^{nl}} 
= 0 ,
\end{equation}
where $\mathbf{\tilde{w} }$ is the vector of the 
expansion coefficients for the local exchange potential 
and  $\mathbf{w}^{nl}$ is the projection in the chosen 
basis set of the non-local HF potential, 
$\hat{V}_{\sigma}^{x,nl} ({\mathbf r}, {\mathbf r}')$. 

In this work our expansion basis set will be 
a scaled form of the occupied-virtual products, 
specifically 
\cite{1999G, 2001CN, 2007KF, 2008KF}
\begin{equation} \label{g}
g_{\mu \sigma} ({\mathbf r}) = 
\frac{ \phi_{a\sigma}({\mathbf r}) \phi_{i\sigma}({\mathbf r}) }
{ \sqrt{ \varepsilon_{a \sigma}- \varepsilon_{i \sigma} } }.
\end{equation} 
Thus, the elements of the vector $\mathbf{w}^{nl}$ 
are 
\begin{equation}\label{nonlocalelements}
w_{ai \sigma}^{nl} = \int d{\bf r} d{\bf r}' 
\frac{ \phi_{a \sigma}({\bf r}) \phi_{i \sigma}({\bf r'}) } %
{\sqrt{\varepsilon_{a \sigma}-\varepsilon_{i \sigma}}}
V^{x,nl}_{\sigma}({\bf r},{\bf r}') ,
\end{equation}
and the matrix elements $M^{\sigma}_{\mu,ia}$ reduce to 
\begin{equation}
M_{ia,jb}^{\sigma} = 
\frac{(\phi_{a \sigma} \phi_{i \sigma} | \phi_{b \sigma} \phi_{j \sigma})} %
{\sqrt{\varepsilon_{a \sigma}-\varepsilon_{i \sigma}} %
\sqrt{\varepsilon_{b \sigma}-\varepsilon_{j \sigma}}} .
\end{equation}

Note that with the basis set of occupied-virtual 
products it is not possible to get any term having 
a $1/r$ asymptotic decay. 
This is corrected by the addition to our basis set 
$\{ g_{\mu \sigma} ({\mathbf r}) \}$
of the Fermi-Amaldi function 
$s({\mathbf r})=\rho({\bf r}) / N$, 
where $N$ is the number of electrons of the system. 
That procedure reproduces the Fermi-Amaldi potential 
for long distances and will make the xOEP HOMO 
eigenenergies to be very close, but not equal, to 
those found using the HF method. 
This prescription \cite{2002YW}
is very different to the one adopted when the exchange 
potential is expanded in an auxiliary basis set 
\cite{2001DSG}.

\section{Computational Details}%
\label{Sec:DetailsResults}

As the products of the occupied and virtual 
states are linearly dependent, the matrix 
$\mathbf{M M}^{\dagger}$ 
appearing in Eq.~(\ref{nablaOEP}) is singular and 
the equation cannot be solved by inversion \cite{2008GHJL}. 
Following the argument given in Refs.~\cite{2010FKF} 
and \cite{1983H}, only the linearly independent 
functions $f_{\mu \sigma}({\bf r})$ 
(or $g_{\mu \sigma}({\bf r})$) 
can be used in the expansion of the xOEP, 
i.~e. for a faithful solution of the xOEP equation, 
Eq.~(\ref{nablaOEP}), linearly-independent orbital 
products must be employed. 
We will then apply the Truncated Singular Value 
decomposition (TSVD) technique to separate the linear 
dependent occupied-virtual products and the independent 
ones, seeking for the linear independent set of 
products by diagonalization of the 
$\mathbf{M M}^{\dagger}$ matrix. 
In order to fulfill this condition, a threshold is chosen 
to discriminate the elements of the 
$\mathbf{M M}^{\dagger}$ matrix that correspond 
to the linearly independent functions 
\footnote{
If all the orbital products were linearly-independent 
(thus making the matrix $\mathbf{M M}^{\dagger}$ perfectly invertible), 
then Eq.~(\ref{nablaOEP}) would have had a unique solution that 
would correspond to the lowest variational energy obtainable 
with the functional given in Eq.~(\ref{xenergy}), that is the HF 
energy $E_{HF}$.
}. 
In this way, the mapping between the density and density matrix 
becomes non-unique and a solution with the energy 
$E_{xOEP} > E_{HF}$ is obtained %
\cite{1983H, 1986H}.

In our case, the TSVD method requires the diagonalization of 
the matrix $\mathbf{M M}^{\dagger}$, in general a very large one. 
But as the STOs represent the unoccupied orbitals in 
a much more efficient way than the GTOs, 
when Slater-type orbital basis sets are used 
the size of the matrix $\mathbf{M M}^{\dagger}$, 
and consequently the expansion space for the local potential 
in Eq.~(\ref{vxspan}), 
is much smaller than when using GTOs. 
This is the main point of this paper: when STO basis sets are 
used, the computational effort 
(in memory size and in speed of the calculations)  
for each self-consistent cycle is much lower, 
whereas the quality of the results is preserved.

The algorithm outlined in the previous Section was implemented 
in the SMILES suite of programs \cite{2008SMILES}, 
which employs the STO basis sets in quantum 
chemical calculations. 
We will compare the results of the xOEP calculations 
obtained with the Slater-type basis functions (xOEP-STO) 
to both numerical exact solutions and xOEP results 
obtained with the use of the Gaussian-type basis 
functions (xOEP-GTO). 
The latter results were obtained with the use of the 
MOLPRO2008.1 code \cite{MOLPRO_brief}, 
where the xOEP formalism employing Gaussian-type basis sets 
was recently implemented by some of us \cite{2007KF, 2008KF}.
In order to use comparable basis sets, the 
correlation-consistent basis sets of Dunning 
(cc-pVTZ, cc-pVQZ, and cc-pV5Z) were used in the 
xOEP-GTO calculations, 
and the STO of similar quality (VB1 for cc-pVTZ, 
VB2 for cc-pVQZ and VB3 for cc-pV5Z) \cite{2008SMILES} 
were selected for the xOEP-STO calculations. 
These STO and GTO basis sets yield the total HF energies 
in close agreement 
(see Tables~\ref{TotalBe} -- \ref{TotalCO} below). 
All the basis sets we have used (xOEP-STO and xOEP-GTO 
calculations) were employed in their uncontracted form.

Due to the few exact numerical xOEP solutions found in the 
literature, 
and to the small number of available STOs for atomic and 
molecular computations, 
the calculations we present here were performed for the 
Be and Ne atoms and for the LiH, BH, Li$_{2}$ and CO 
molecules. 
The numerical solutions were given in Makmal et al. \cite{2009MKK} 
and we have used the same internuclear distances (in a.u.):
$3.015$ for LiH, $2.336$ for BH, $5.051$ for Li$_{2}$ and 
$2.132$ for CO. 
This distance for the CO molecule has also been used by 
He{\ss}elmann et al. \cite{2007HGDSG} in the xOEP-GTO 
solution, and we will also compare our results with 
theirs in Section \ref{Sec:Results}.

\section{Results}%
\label{Sec:Results}

The dependence of the xOEP-STO and the xOEP-GTO total energies  
on the size of the basis set and on the TSVD cutoff criterion 
$\varepsilon$ for neglecting (near) zero eigenvalues of the 
matrix $\mathbf{M M}^{\dagger}$ has been investigated.
The results are collected in Tables 
\ref{TotalBe}, \ref{TotalLiH} and \ref{TotalCO}. 
In the calculations, the TSVD cutoff criterion $\varepsilon$ 
was varied in the range $10^{-2}-10^{-6}$. 
Note that when $\varepsilon = 10^{-2}$ is employed, the expansion 
space for the potential is very small because only a few 
eigenvalues of the matrix $\mathbf{M M}^{\dagger}$ are greater 
than $\varepsilon$, yielding energies noticeably above the 
numeric xOEP values. 
For that reason, we have not reported that energies 
in the tables.
Reducing the TSVD cutoff $\varepsilon$ makes the 
expansion space bigger and, as a consequence, the total 
xOEP-STO and xOEP-GTO energies decrease, approaching to the 
accurate numeric xOEP values. 
For $\varepsilon$ in the range $10^{-3}-10^{-5}$, the total 
xOEP energies remain constant to within a fraction of mHa. 
When the xOEP-GTO method is used with a very tight cutoff 
criteria ($\varepsilon \leq 10^{-6}$), the iterative solution 
of the xOEP equation (\ref{nablaOEP}) becomes unstable and 
the xOEP total energy collapses towards the HF energy. 
As a matter of fact, the procedure breaks-down as the matrix 
$\mathbf{M M}^{\dagger}$ becomes noninvertible. 
On the other hand, the xOEP-STO implementation shows somewhat 
greater stability and begins to break down at smaller values 
of $\varepsilon$, i.~e. when 
$\varepsilon \leq 10^{-7}$. 
This can be attributed to the fact that, with the use of 
STO functions, the expansion set of the potential is much 
smaller than with the more traditional Gaussian-basis sets 
(see below).

The number of the eigenfunctions of the matrix 
$\mathbf{M M}^{\dagger}$ used for the expansion 
of the potential is also given in Tables 
\ref{TotalBe}, \ref{TotalLiH} and \ref{TotalCO}, 
as well as the total dimension of the matrix. 
It is seen that the dimension of $\mathbf{M M}^{\dagger}$
is considerably smaller in the xOEP-STO method, and the 
dimension of the potential expansion space does not 
grow as fast as in the case of the xOEP-GTO method. 
So, the xOEP-STO implementation gives a noticeably 
memory savings and a greater stability with respect 
to the cutoff criterion of the TSVD procedure.

For all the systems in Tables 
\ref{TotalBe}, \ref{TotalLiH} and \ref{TotalCO}, 
the differences between the xOEP-STO and the xOEP-GTO 
total energies are typically smaller than 1 mHa. 
When large basis sets are used the total xOEP-STO 
and xOEP-GTO energies approach the exact numeric 
values with an accuracy better than 100 $\mu$Ha. 
It is important to stress here that the xOEP and 
HF total energies converge in a somewhat different 
way to respect to the basis set size (remember that 
the basis sets we use in this paper were nor developed neither 
optimized for the xOEP calculations) so the differences 
$E_{xOEP} - E_{HF}$ oscillate.

Table \ref{TotalxOEP} summarizes the results of calculations 
for Be and Ne atoms and a number of diatomic molecules studied 
by Makmal et al. \cite{2009MKK} using the real space xOEP 
method. 
For the sake of comparison, the xOEP energies obtained 
by He{\ss}elmann et al. \cite{2007HGDSG} with the use 
of balanced auxiliary basis sets for the potential are 
also shown when available. 
Note that the xOEP-STO energies are typically in a 
somewhat better agreement with the numerical values 
than the xOEP-GTO energies obtained with similar 
basis set 
(the STO basis sets give results about $0.1$ mHa below 
the energies obtained by He{\ss}elmann et al.).

Table \ref{OrbitalxOEP} collects the energies of the occupied 
orbitals obtained with the use of the xOEP-STO and xOEP-GTO 
methods for the Be atoms and the LiH and Li$_{2}$ molecules.
There is a good agreement between the numerical values of these 
orbital energies and the energies obtained with the two xOEP 
methods. 
Furthermore, the orbital energies from the xOEP-STO and from 
the xOEP-GTO calculations agree with each other to within a 
few mHa.

Besides the total and orbital xOEP energies, we have also 
studied the energy decomposition into the kinetic, 
nuclear-electron attraction, electron-electron repulsion 
and the exchange energies.
Table \ref{Edecomp} presents the above components of the 
total xOEP energy as obtained using the STO and GTO basis 
sets for the 
Be atom and the CO molecule (there are no numerical 
solutions available in this case for the xOEP). 

\begin{figure}[h]
\begin{center}
\includegraphics[width=12cm]{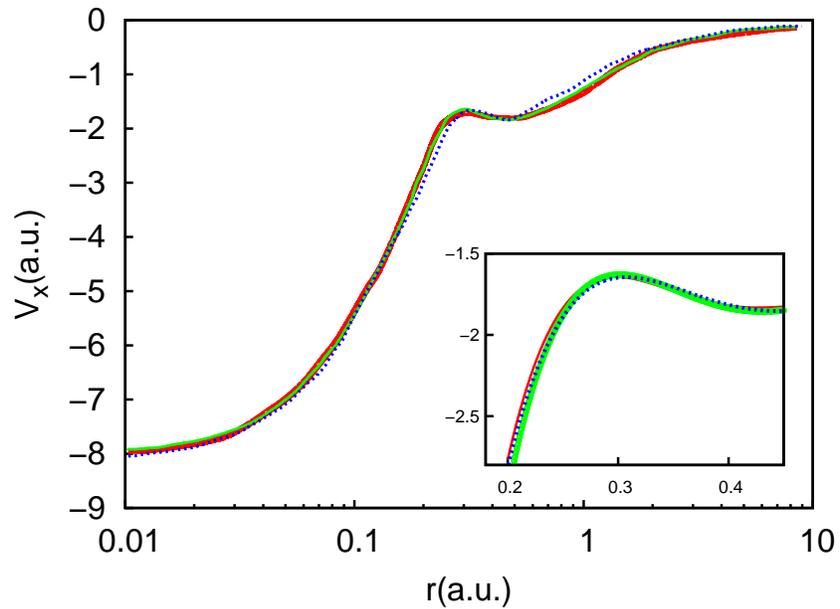}
\end{center}
\vspace*{-1.cm}
\caption{xOEP potential for the Ne atom. 
Atomic units are used in both coordinate axis 
as well as a logarithmic scale in the OX axis.
The dashed blue line corresponds to the xOEP-GTO 
calculation of Fern\'{a}ndez et al. \cite{2010FKF}, 
the red line represents the results of this paper and 
the green one reflects the numerical results of 
Kurth and Pittalis \cite{2006KP}.
In the inset (where the OX scale is now linear) 
we show in more detail the region around the shoulder.
}
\label{Fig:Ne}
\end{figure}

\begin{figure}[h]
\begin{center}
\includegraphics[width=12cm]{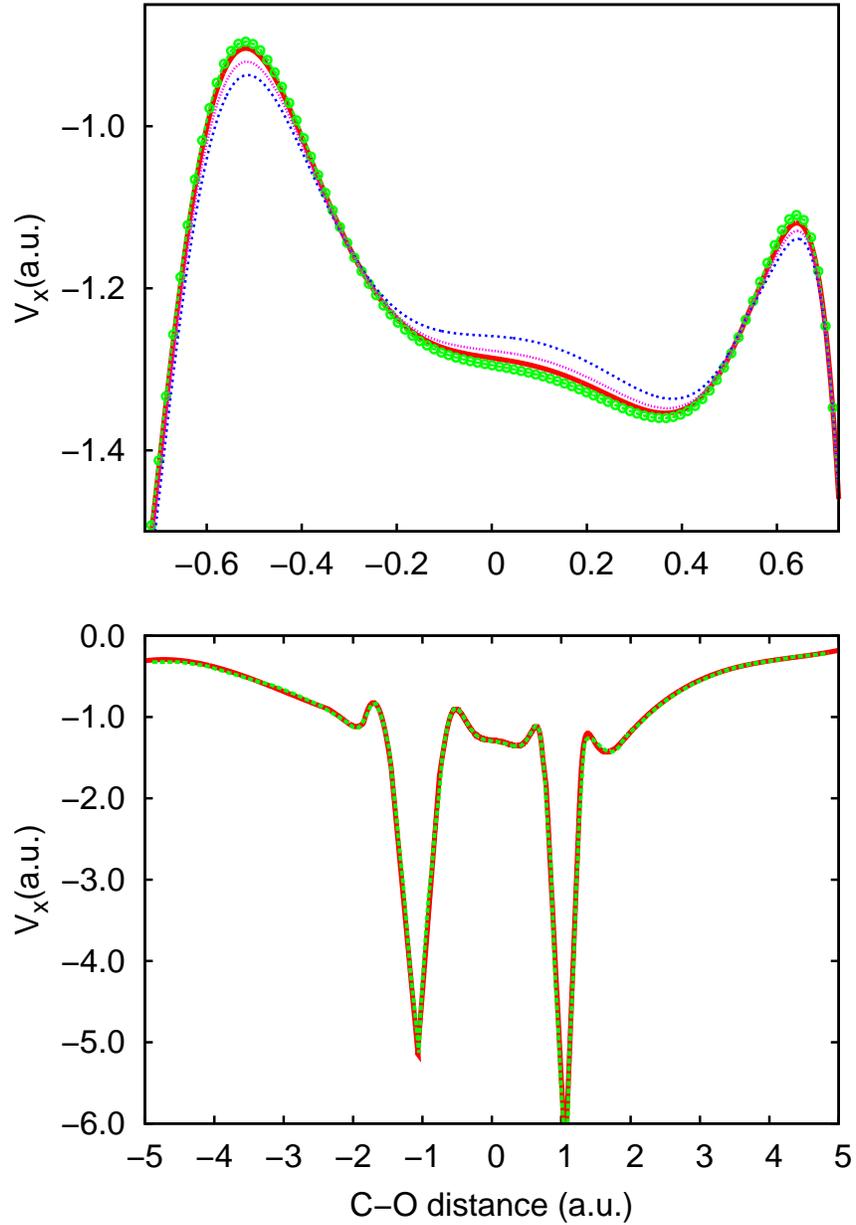}
\end{center}
\vspace*{-1.cm}
\caption{xOEP potential for the CO molecule along the axis 
of the molecule. 
Atomic units are used in both coordinate axis 
(we have used an internuclear distance of $2.132$ a.u.).
In the lower panel, the red line show the results of 
this paper and the green one reflects the xOEP-GTO 
results of He{\ss}elmann et al. \cite{2007HGDSG}.
In the upper panel the region between the nuclei of the 
molecule is depicted in more detail;
the xOEP-GTO results are shown again in green, 
and results for xOEP-STO calculations with 
different STO basis sets \cite{2008SMILES} 
are also plotted
(VB1, blue dashed line;
VB2, magenta dotted line; VB3, red thick line).
}
\label{Fig:CO}
\end{figure}

Figs.~\ref{Fig:Ne} and \ref{Fig:CO} show the results of 
the xOEP potential for the Ne atom and the CO molecule  
along the main axis. 
For the Ne atom, note the close agreement between our result  
and that obtained with the exact numerical calculation by 
Kurth and Pittalis \cite{2006KP}, 
thus yielding a smooth potential that shows only small 
deviations from the numerical potential; 
in any case, these xOEP potentials are also very similar to 
that evaluated with a Gaussian basis set \cite{2010FKF} 
using the procedure presented in Refs.~\cite{2007KF}
and \cite{2008KF}.
In the lower panel of Fig.~\ref{Fig:CO} the results we 
have obtained using the VB3 basis set for the CO molecule 
are compared with the calculation by 
He{\ss}elmann et al. \cite{2007HGDSG}, 
using an auxiliary basis set within a Gaussian 
representation; 
the STO results show a good agreement with the 
xOEP-GTO potential. 
For the sake of completeness, in the upper panel we also 
present several other calculations for the internuclear 
region, using STO basis sets of different quality 
(VB1, VB2 and VB3).
Note that our xOEP potentials do not present any  
unphysical wiggle as those found by Staronerov et al. 
\cite{2006SSD} 
and have a good agreement with both the numerical 
calculation and the xOEP-GTO solution by Hessellmann et al. 
\cite{2007HGDSG} 
with the use of auxiliary basis set.

In summary, the previous results show that both xOEP-STO and 
xOEP-GTO methods yield close results to the numerical exact 
ones for the total energies, the one-electron energies 
of the occupied orbitals and the xOEP potentials.

\section{Conclusions and discussion}
\noindent
The implementation of the xOEP formalism with 
Slater-type basis functions has been developed. 
This procedure has been tested in a small group 
of closed shell atoms and diatomic molecules, for which 
both numerical xOEP solutions and STO basis sets 
are available.  
When compared to the exact numerical solution of the 
xOEP equations we have obtained very good results; 
they are even a bit better than those given by the 
Gaussian-type basis sets procedure. 
On the other hand, both xOEP-STO and xOEP-GTO 
results obtained with the prescription proposed 
in Refs.~[\onlinecite{2007KF}] and [\onlinecite{2008KF}] 
give energies about $0.1$ mHa below 
those obtained for xOEP-GTO by He{\ss}elmann et al., 
and thus they are closer to the exact results \cite{2007HGDSG}.

The new method leads to a considerably more compact 
expansion space for the xOEP local multiplicative 
potential, yielding noticeable savings in the computational 
effort to be done in each one of the cycles of the 
self-consistent procedure.  
Yet another advantage of using the Slate-type basis sets is 
that, within the TSVD algorithm, fewer eigenvalues of the 
(near) singular matrix $\mathbf{M M}^{\dagger}$ need to be 
employed, which leads to an increased numeric stability of 
the xOEP-STO method as compared to the xOEP-GTO algorithm. 

As a final remark, it is known that a more efficient xOEP 
algorithm can be developed based on the use of the incomplete 
Cholesky decomposition technique.
The use of this technique would facilitate the application 
of the xOEP-STO method to larger molecules. 
This implementation is currently in progress and will be reported 
elsewhere. 

On the other hand, this work is a first step to develop a local 
potential formalism for both exchange and correlation. 
Due to the smaller memory requirements of the STO scheme we have 
presented here, it can be used to study bigger molecules 
than those that can be solved with the standard GTO approach. 
The implementation of the correlation part of the potential 
is under development.

\section{Acknowledgments}
The authors thank Prof. Jaime Fern\'{a}ndez Rico 
and his research group for providing a copy of the SMILES suite 
of programs and supporting this work. 
The authors acknowledge the financial support from the Spanish Ministerio 
de Ciencia e Innovaci\'{o}n through the research Project Grant 
FIS2010-21282-C02-02.


\begin{thebibliography}{36}
\expandafter\ifx\csname natexlab\endcsname\relax\def\natexlab#1{#1}\fi
\expandafter\ifx\csname bibnamefont\endcsname\relax
  \def\bibnamefont#1{#1}\fi
\expandafter\ifx\csname bibfnamefont\endcsname\relax
  \def\bibfnamefont#1{#1}\fi
\expandafter\ifx\csname citenamefont\endcsname\relax
  \def\citenamefont#1{#1}\fi
\expandafter\ifx\csname url\endcsname\relax
  \def\url#1{\texttt{#1}}\fi
\expandafter\ifx\csname urlprefix\endcsname\relax\def\urlprefix{URL }\fi
\providecommand{\bibinfo}[2]{#2}
\providecommand{\eprint}[2][]{\url{#2}}

\bibitem[{\citenamefont{Hohenberg and Kohn}(1964)}]{1964HK}
\bibinfo{author}{\bibfnamefont{P.}~\bibnamefont{Hohenberg}} \bibnamefont{and}
  \bibinfo{author}{\bibfnamefont{W.}~\bibnamefont{Kohn}},
  \bibinfo{journal}{Phys. Rev. B} \textbf{\bibinfo{volume}{136}},
  \bibinfo{pages}{864} (\bibinfo{year}{1964}).

\bibitem[{\citenamefont{Kohn and Sham}(1965)}]{1965KS}
\bibinfo{author}{\bibfnamefont{W.}~\bibnamefont{Kohn}} \bibnamefont{and}
  \bibinfo{author}{\bibfnamefont{L.~J.} \bibnamefont{Sham}},
  \bibinfo{journal}{Phys. Rev. A} \textbf{\bibinfo{volume}{140}},
  \bibinfo{pages}{1133} (\bibinfo{year}{1965}).

\bibitem[{\citenamefont{Kummel and Kronik}(2008)}]{2008KK}
\bibinfo{author}{\bibfnamefont{S.}~\bibnamefont{Kummel}} \bibnamefont{and}
  \bibinfo{author}{\bibfnamefont{L.}~\bibnamefont{Kronik}},
  \bibinfo{journal}{Rev. Mod. Phys.} \textbf{\bibinfo{volume}{80}},
  \bibinfo{pages}{3} (\bibinfo{year}{2008}).

\bibitem[{\citenamefont{Sharp and Horton}(1053)}]{1953SH}
\bibinfo{author}{\bibfnamefont{R.~T.} \bibnamefont{Sharp}} \bibnamefont{and}
  \bibinfo{author}{\bibfnamefont{G.~K.} \bibnamefont{Horton}},
  \bibinfo{journal}{Phys. Rev} \textbf{\bibinfo{volume}{90}},
  \bibinfo{pages}{317} (\bibinfo{year}{1053}).

\bibitem[{\citenamefont{Talman and Shadwick}(1976)}]{1976TS}
\bibinfo{author}{\bibfnamefont{J.~D.} \bibnamefont{Talman}} \bibnamefont{and}
  \bibinfo{author}{\bibfnamefont{W.~F.} \bibnamefont{Shadwick}},
  \bibinfo{journal}{Phys. Rev. A} \textbf{\bibinfo{volume}{14}},
  \bibinfo{pages}{36} (\bibinfo{year}{1976}).

\bibitem[{\citenamefont{Engel and Dreizler}(1999)}]{1999ED}
\bibinfo{author}{\bibfnamefont{E.}~\bibnamefont{Engel}} \bibnamefont{and}
  \bibinfo{author}{\bibfnamefont{R.~M.} \bibnamefont{Dreizler}},
  \bibinfo{journal}{J. Comp. Chem.} \textbf{\bibinfo{volume}{20}},
  \bibinfo{pages}{31} (\bibinfo{year}{1999}).

\bibitem[{\citenamefont{Sanhi et~al.}(1982)\citenamefont{Sanhi, Gruenebaum, and
  Perdew}}]{1982SGP}
\bibinfo{author}{\bibfnamefont{V.}~\bibnamefont{Sanhi}},
  \bibinfo{author}{\bibfnamefont{J.}~\bibnamefont{Gruenebaum}},
  \bibnamefont{and} \bibinfo{author}{\bibfnamefont{J.~P.}
  \bibnamefont{Perdew}}, \bibinfo{journal}{Phys. Rev. B}
  \textbf{\bibinfo{volume}{26}}, \bibinfo{pages}{4371} (\bibinfo{year}{1982}).

\bibitem[{\citenamefont{Engel and Vosko}(1993)}]{1993EV}
\bibinfo{author}{\bibfnamefont{E.}~\bibnamefont{Engel}} \bibnamefont{and}
  \bibinfo{author}{\bibfnamefont{S.~H.} \bibnamefont{Vosko}},
  \bibinfo{journal}{Phys. Rev. A} \textbf{\bibinfo{volume}{47}},
  \bibinfo{pages}{2800} (\bibinfo{year}{1993}).

\bibitem[{\citenamefont{Makmal et~al.}(2009{\natexlab{a}})\citenamefont{Makmal,
  K\"{u}mmel, and L.}}]{2009MKK}
\bibinfo{author}{\bibfnamefont{A.}~\bibnamefont{Makmal}},
  \bibinfo{author}{\bibfnamefont{S.}~\bibnamefont{K\"{u}mmel}},
  \bibnamefont{and} \bibinfo{author}{\bibfnamefont{K.}~\bibnamefont{L.}},
  \bibinfo{journal}{J. Chem. Theory Comput.} \textbf{\bibinfo{volume}{5}},
  \bibinfo{pages}{1731} (\bibinfo{year}{2009}{\natexlab{a}}).

\bibitem[{\citenamefont{Staedele et~al.}(1999)\citenamefont{Staedele, Moukara,
  Majewski, Vogl, and G\"{o}rling}}]{1999SMMVG}
\bibinfo{author}{\bibfnamefont{M.}~\bibnamefont{Staedele}},
  \bibinfo{author}{\bibfnamefont{M.}~\bibnamefont{Moukara}},
  \bibinfo{author}{\bibfnamefont{J.~A.} \bibnamefont{Majewski}},
  \bibinfo{author}{\bibfnamefont{P.}~\bibnamefont{Vogl}}, \bibnamefont{and}
  \bibinfo{author}{\bibfnamefont{A.}~\bibnamefont{G\"{o}rling}},
  \bibinfo{journal}{Phys Rev B} \textbf{\bibinfo{volume}{59}},
  \bibinfo{pages}{10031} (\bibinfo{year}{1999}).

\bibitem[{\citenamefont{K\"{o}rdorfer et~al.}(2008)\citenamefont{K\"{o}rdorfer,
  K\"{u}mmel, and M.}}]{2008KKM}
\bibinfo{author}{\bibfnamefont{T.}~\bibnamefont{K\"{o}rdorfer}},
  \bibinfo{author}{\bibfnamefont{S.}~\bibnamefont{K\"{u}mmel}},
  \bibnamefont{and} \bibinfo{author}{\bibfnamefont{M.}~\bibnamefont{M.}},
  \bibinfo{journal}{J. Chem. Phys} \textbf{\bibinfo{volume}{129}},
  \bibinfo{pages}{014110} (\bibinfo{year}{2008}).

\bibitem[{\citenamefont{Engel and Schmid}(2009)}]{2009ES}
\bibinfo{author}{\bibfnamefont{E.}~\bibnamefont{Engel}} \bibnamefont{and}
  \bibinfo{author}{\bibfnamefont{R.~N.} \bibnamefont{Schmid}},
  \bibinfo{journal}{Phys Rev Lett} \textbf{\bibinfo{volume}{103}},
  \bibinfo{pages}{036404} (\bibinfo{year}{2009}).

\bibitem[{\citenamefont{Makmal et~al.}(2009{\natexlab{b}})\citenamefont{Makmal,
  Armiento, Engel, Kronik, and K\"{u}mmel}}]{2009MAEKK}
\bibinfo{author}{\bibfnamefont{A.}~\bibnamefont{Makmal}},
  \bibinfo{author}{\bibfnamefont{R.}~\bibnamefont{Armiento}},
  \bibinfo{author}{\bibfnamefont{E.}~\bibnamefont{Engel}},
  \bibinfo{author}{\bibfnamefont{L.}~\bibnamefont{Kronik}}, \bibnamefont{and}
  \bibinfo{author}{\bibfnamefont{S.}~\bibnamefont{K\"{u}mmel}},
  \bibinfo{journal}{Phys Rev B} \textbf{\bibinfo{volume}{80}},
  \bibinfo{pages}{161204(R)} (\bibinfo{year}{2009}{\natexlab{b}}).

\bibitem[{\citenamefont{Engel}(2009)}]{2009Engel}
\bibinfo{author}{\bibfnamefont{E.}~\bibnamefont{Engel}}, \bibinfo{journal}{Phys
  Rev B} \textbf{\bibinfo{volume}{80}}, \bibinfo{pages}{161205(R)}
  (\bibinfo{year}{2009}).

\bibitem[{\citenamefont{Krieger et~al.}(1992)\citenamefont{Krieger, Li, and
  Iafrate}}]{1992KLI}
\bibinfo{author}{\bibfnamefont{J.~B.} \bibnamefont{Krieger}},
  \bibinfo{author}{\bibfnamefont{Y.}~\bibnamefont{Li}}, \bibnamefont{and}
  \bibinfo{author}{\bibfnamefont{G.~J.} \bibnamefont{Iafrate}},
  \bibinfo{journal}{Phys Rev A} \textbf{\bibinfo{volume}{46}},
  \bibinfo{pages}{5453} (\bibinfo{year}{1992}).

\bibitem[{\citenamefont{Li et~al.}(1993)\citenamefont{Li, Krieger, and
  Iafrate}}]{1993LKI}
\bibinfo{author}{\bibfnamefont{Y.}~\bibnamefont{Li}},
  \bibinfo{author}{\bibfnamefont{J.~B.} \bibnamefont{Krieger}},
  \bibnamefont{and} \bibinfo{author}{\bibfnamefont{G.~J.}
  \bibnamefont{Iafrate}}, \bibinfo{journal}{Phys Rev A}
  \textbf{\bibinfo{volume}{47}}, \bibinfo{pages}{165} (\bibinfo{year}{1993}).

\bibitem[{\citenamefont{Della~Sala and G\"{o}rling}(2001)}]{2001DSG}
\bibinfo{author}{\bibfnamefont{F.}~\bibnamefont{Della~Sala}} \bibnamefont{and}
  \bibinfo{author}{\bibfnamefont{A.}~\bibnamefont{G\"{o}rling}},
  \bibinfo{journal}{J. Chem. Phys} \textbf{\bibinfo{volume}{115}},
  \bibinfo{pages}{5718} (\bibinfo{year}{2001}).

\bibitem[{\citenamefont{Staroverov et~al.}(2006)\citenamefont{Staroverov,
  Scuseria, and Davidson}}]{2006SSD}
\bibinfo{author}{\bibfnamefont{V.~N.} \bibnamefont{Staroverov}},
  \bibinfo{author}{\bibfnamefont{G.~E.} \bibnamefont{Scuseria}},
  \bibnamefont{and} \bibinfo{author}{\bibfnamefont{E.~R.}
  \bibnamefont{Davidson}}, \bibinfo{journal}{J. Chem. Phys}
  \textbf{\bibinfo{volume}{125}}, \bibinfo{pages}{081104}
  (\bibinfo{year}{2006}).

\bibitem[{\citenamefont{Ivanov et~al.}(1999)\citenamefont{Ivanov, Hirata, and
  Bartlett}}]{1999IHB}
\bibinfo{author}{\bibfnamefont{S.}~\bibnamefont{Ivanov}},
  \bibinfo{author}{\bibfnamefont{S.}~\bibnamefont{Hirata}}, \bibnamefont{and}
  \bibinfo{author}{\bibfnamefont{R.~J.} \bibnamefont{Bartlett}},
  \bibinfo{journal}{Phys Rev Lett} \textbf{\bibinfo{volume}{83}},
  \bibinfo{pages}{5455} (\bibinfo{year}{1999}).

\bibitem[{\citenamefont{Kollmar and Filatov}(2007)}]{2007KF}
\bibinfo{author}{\bibfnamefont{C.}~\bibnamefont{Kollmar}} \bibnamefont{and}
  \bibinfo{author}{\bibfnamefont{M.}~\bibnamefont{Filatov}},
  \bibinfo{journal}{J. Chem. Phys} \textbf{\bibinfo{volume}{127}},
  \bibinfo{pages}{114104} (\bibinfo{year}{2007}).

\bibitem[{\citenamefont{Kollmar and Filatov}(2008)}]{2008KF}
\bibinfo{author}{\bibfnamefont{C.}~\bibnamefont{Kollmar}} \bibnamefont{and}
  \bibinfo{author}{\bibfnamefont{M.}~\bibnamefont{Filatov}},
  \bibinfo{journal}{J. Chem. Phys} \textbf{\bibinfo{volume}{128}},
  \bibinfo{pages}{064101} (\bibinfo{year}{2008}).

\bibitem[{\citenamefont{Hesselman et~al.}(2007)\citenamefont{Hesselman,
  G\"{o}tz, Della~Sala, and G\"{o}rling}}]{2007HGDSG}
\bibinfo{author}{\bibfnamefont{A.}~\bibnamefont{Hesselman}},
  \bibinfo{author}{\bibfnamefont{A.}~\bibnamefont{G\"{o}tz}},
  \bibinfo{author}{\bibfnamefont{F.}~\bibnamefont{Della~Sala}},
  \bibnamefont{and}
  \bibinfo{author}{\bibfnamefont{A.}~\bibnamefont{G\"{o}rling}},
  \bibinfo{journal}{J. Chem. Phys} \textbf{\bibinfo{volume}{127}},
  \bibinfo{pages}{054102} (\bibinfo{year}{2007}).

\bibitem[{\citenamefont{G\"{o}rling}(1999)}]{1999G}
\bibinfo{author}{\bibfnamefont{A.}~\bibnamefont{G\"{o}rling}},
  \bibinfo{journal}{Phys Rev Lett} \textbf{\bibinfo{volume}{83}},
  \bibinfo{pages}{5459} (\bibinfo{year}{1999}).

\bibitem[{\citenamefont{Yang and Wu}(2002)}]{2002YW}
\bibinfo{author}{\bibfnamefont{W.}~\bibnamefont{Yang}} \bibnamefont{and}
  \bibinfo{author}{\bibfnamefont{Q.}~\bibnamefont{Wu}}, \bibinfo{journal}{Phys
  Rev Lett} \textbf{\bibinfo{volume}{89}}, \bibinfo{pages}{143002}
  (\bibinfo{year}{2002}).

\bibitem[{\citenamefont{G\"{o}rling et~al.}(2008)\citenamefont{G\"{o}rling,
  Hesselman, Jones, and Levy}}]{2008GHJL}
\bibinfo{author}{\bibfnamefont{A.}~\bibnamefont{G\"{o}rling}},
  \bibinfo{author}{\bibfnamefont{A.}~\bibnamefont{Hesselman}},
  \bibinfo{author}{\bibfnamefont{M.}~\bibnamefont{Jones}}, \bibnamefont{and}
  \bibinfo{author}{\bibfnamefont{M.}~\bibnamefont{Levy}}, \bibinfo{journal}{J.
  Chem. Phys} \textbf{\bibinfo{volume}{128}}, \bibinfo{pages}{104104}
  (\bibinfo{year}{2008}).

\bibitem[{\citenamefont{Colle and Nesbet}(2001)}]{2001CN}
\bibinfo{author}{\bibfnamefont{R.}~\bibnamefont{Colle}} \bibnamefont{and}
  \bibinfo{author}{\bibfnamefont{R.~K.} \bibnamefont{Nesbet}},
  \bibinfo{journal}{J. Phys. B} \textbf{\bibinfo{volume}{34}},
  \bibinfo{pages}{2475} (\bibinfo{year}{2001}).

\bibitem[{\citenamefont{Harriman}(1983)}]{1983H}
\bibinfo{author}{\bibfnamefont{J.~E.} \bibnamefont{Harriman}},
  \bibinfo{journal}{Phys Rev A} \textbf{\bibinfo{volume}{27}},
  \bibinfo{pages}{632} (\bibinfo{year}{1983}).

\bibitem[{\citenamefont{Fern\'{a}ndez~Rico
  et~al.}(2008)\citenamefont{Fern\'{a}ndez~Rico, Ema, L\'{o}pez, Ram\'{i}rez,
  and Ishida}}]{2008SMILES}
\bibinfo{author}{\bibfnamefont{J.}~\bibnamefont{Fern\'{a}ndez~Rico}},
  \bibinfo{author}{\bibfnamefont{I.}~\bibnamefont{Ema}},
  \bibinfo{author}{\bibfnamefont{R.}~\bibnamefont{L\'{o}pez}},
  \bibinfo{author}{\bibfnamefont{G.}~\bibnamefont{Ram\'{i}rez}},
  \bibnamefont{and} \bibinfo{author}{\bibfnamefont{K.}~\bibnamefont{Ishida}},
  \emph{\bibinfo{title}{Recent Advances in Computational Chemistry: Molecular
  Integrals over Slater Orbitals}} (\bibinfo{publisher}{Transworld Research
  Network, 978-81-7895-370-0}, \bibinfo{year}{2008}), chap.
  \bibinfo{chapter}{SMILES: A package for molecular calculations with STO
  software, third generation}.

\bibitem[{\citenamefont{Heaton-Burguess
  et~al.}(2007)\citenamefont{Heaton-Burguess, Bulat, and Yang}}]{2007HBY}
\bibinfo{author}{\bibfnamefont{T.}~\bibnamefont{Heaton-Burguess}},
  \bibinfo{author}{\bibfnamefont{F.~A.} \bibnamefont{Bulat}}, \bibnamefont{and}
  \bibinfo{author}{\bibfnamefont{W.}~\bibnamefont{Yang}},
  \bibinfo{journal}{Phys. Rev. Lett.} \textbf{\bibinfo{volume}{98}},
  \bibinfo{pages}{256401} (\bibinfo{year}{2007}).

\bibitem[{\citenamefont{Ivanov et~al.}(2002)\citenamefont{Ivanov, Hirata, and
  Bartlett}}]{2002IHB}
\bibinfo{author}{\bibfnamefont{S.}~\bibnamefont{Ivanov}},
  \bibinfo{author}{\bibfnamefont{S.}~\bibnamefont{Hirata}}, \bibnamefont{and}
  \bibinfo{author}{\bibfnamefont{R.~J.} \bibnamefont{Bartlett}},
  \bibinfo{journal}{J. Chem. Phys} \textbf{\bibinfo{volume}{116}},
  \bibinfo{pages}{1269} (\bibinfo{year}{2002}).

\bibitem[{\citenamefont{Fern\'{a}ndez et~al.}(2010)\citenamefont{Fern\'{a}ndez,
  Kollmar, and Filatov}}]{2010FKF}
\bibinfo{author}{\bibfnamefont{J.~J.} \bibnamefont{Fern\'{a}ndez}},
  \bibinfo{author}{\bibfnamefont{C.}~\bibnamefont{Kollmar}}, \bibnamefont{and}
  \bibinfo{author}{\bibfnamefont{M.}~\bibnamefont{Filatov}},
  \bibinfo{journal}{Phys. Rev. A} \textbf{\bibinfo{volume}{82}},
  \bibinfo{pages}{0225081} (\bibinfo{year}{2010}).

\bibitem[{\citenamefont{Harriman}(1986)}]{1986H}
\bibinfo{author}{\bibfnamefont{J.~E.} \bibnamefont{Harriman}},
  \bibinfo{journal}{Phys Rev A} \textbf{\bibinfo{volume}{34}},
  \bibinfo{pages}{29} (\bibinfo{year}{1986}).

\bibitem[{\citenamefont{Werner et~al.}(2010)\citenamefont{Werner, Knowles,
  Knizia, Manby, {Sch\"{u}tz} et~al.}}]{MOLPRO_brief}
\bibinfo{author}{\bibfnamefont{H.-J.} \bibnamefont{Werner}},
  \bibinfo{author}{\bibfnamefont{P.~J.} \bibnamefont{Knowles}},
  \bibinfo{author}{\bibfnamefont{G.}~\bibnamefont{Knizia}},
  \bibinfo{author}{\bibfnamefont{F.~R.} \bibnamefont{Manby}},
  \bibinfo{author}{\bibfnamefont{M.}~\bibnamefont{{Sch\"{u}tz}}},
  \bibnamefont{et~al.}, \emph{\bibinfo{title}{Molpro, version 2010.1, a package
  of ab initio programs}} (\bibinfo{year}{2010}).

\bibitem[{\citenamefont{Kurth and Pittalis}(2006)}]{2006KP}
\bibinfo{author}{\bibfnamefont{S.}~\bibnamefont{Kurth}} \bibnamefont{and}
  \bibinfo{author}{\bibfnamefont{S.}~\bibnamefont{Pittalis}},
  \emph{\bibinfo{title}{Computational Nanoscience: Do It Yourself!}}
  (\bibinfo{publisher}{John von Neumann Institute for Computing, Julich, ISBN
  3-00-017350-1}, \bibinfo{year}{2006}), vol.~\bibinfo{volume}{31} of
  \emph{\bibinfo{series}{NIC Series}}, chap. \bibinfo{chapter}{\textit{The
  Optimized Effective Potential Method and LDA + U}}, pp.
  \bibinfo{pages}{299--334}.

\bibitem[{\citenamefont{Bunge et~al.}(1992)\citenamefont{Bunge, Barrientos,
  Bunge, and Cogordan}}]{1992BBBC}
\bibinfo{author}{\bibfnamefont{C.~F.} \bibnamefont{Bunge}},
  \bibinfo{author}{\bibfnamefont{J.~A.} \bibnamefont{Barrientos}},
  \bibinfo{author}{\bibfnamefont{A.~V.} \bibnamefont{Bunge}}, \bibnamefont{and}
  \bibinfo{author}{\bibfnamefont{J.~A.} \bibnamefont{Cogordan}},
  \bibinfo{journal}{Phys Rev A} \textbf{\bibinfo{volume}{46}},
  \bibinfo{pages}{3691} (\bibinfo{year}{1992}).

\bibitem[{\citenamefont{Jensen}(2005)}]{2005Jensen}
\bibinfo{author}{\bibfnamefont{F.}~\bibnamefont{Jensen}},
  \bibinfo{journal}{Theor. Chem. Acc.} \textbf{\bibinfo{volume}{113}},
  \bibinfo{pages}{187} (\bibinfo{year}{2005}).

\end{thebibliography}



\begin{table}
\centering

\caption{Total energies (Hartree) for the Be atom using the HF 
and the xOEP methods with the STO and the GTO basis sets.  
In the first column we indicate the value $\varepsilon$ 
used as a threshold in the TSVD decomposition of the 
TSVD decomposition of the $\mathbf{M M}^{\dagger}$ 
matrix in the xOEP method.  
For the STO and GTO xOEP results we have included, within 
brackets, the total number of occupied-virtual products
(first number in the brackets) and the number of them that 
are used (second number in the brackets) in each calculation. }
\label{TotalBe}

\begin{tabular}{|c|c|c|c|c|}
\hline
\hline
$\varepsilon$ & STO--basis & $E_{xOEP}^{STO}$ & GTO--basis & $E_{xOEP}^{GTO}$ \\
\hline
$10^{-3}$     & VB1        & -14.57238 \ (12/3)  & cc-pVTZ    & -14.57233 \ (26/10) \\
              & VB2        & -14.57240 \ (17/4)  & cc-pVQZ     & -14.57234 \ (38/17) \\
              & VB3        & -14.57240 \ (22/5)  & cc-pV5Z     & -14.57227 \ (52/20) \\
\hline
$10^{-4}$     & VB1        & -14.57242 \ (12/4)  & cc-pVTZ    & -14.57233 \ (26/14) \\
              & VB2        & -14.57246 \ (17/5)  & cc-pVQZ    & -14.57256 \ (38/20)  \\
              & VB3        & -14.57245 \ (22/7)  & cc-pV5Z    & -14.57255 \ (52/26) \\
\hline
$10^{-5}$     & VB1        & -14.57244 \ (12/6)  & cc-pVTZ    & -14.57241 \ (26/16) \\
              & VB2        & -14.57256 \ (17/8)  & cc-pVQZ    & -14.57256 \ (38/21) \\
              & VB3        & -14.57254 \ (22/13) & cc-pV5Z    & -14.57255 \ (52/30) \\
\hline
\hline
              & STO--basis & $E_{HF}^{STO}$ & GTO--basis    & $E_{HF}^{GTO}$ \\ 
\hline
              & VB1        & -14.57297      & cc-pVTZ      & -14.57287 \\ 
HF            & VB2        & -14.57298      & cc-pVQZ      & -14.57296 \\
              & VB3        & -14.57301      & cc-pV5Z      & -14.57301 \\
\hline
\hline
\end{tabular}

\end{table}

\newpage

\begin{table}
\centering

\caption{Total energies (Hartree) for the LiH molecule using the HF 
and the xOEP methods with the STO and the GTO basis sets.
In the first column we indicate the value $\varepsilon$ used as a 
threshold in the TSVD decomposition of the $\mathbf{M M}^{\dagger}$ 
matrix in the xOEP method.
For the xOEP-STO and xOEP-GTO results we have included, within 
brackets, the total number of occupied-virtual products 
(first number within the brackets) 
and the number of them that are used (second number within the 
brackets) in each calculation. }
\label{TotalLiH}

\begin{tabular}{|c|c|c|c|c|}
\hline
\hline
$\varepsilon$ & STO--basis & $E_{xOEP}^{STO}$ & GTO--basis & $E_{xOEP}^{GTO}$ \\
\hline
$10^{-3}$     & VB1        & -7.98651  \ (14/3)  & cc-pVTZ    & -7.98660  \ (58/24) \\
              & VB2        & -7.98660  \ (18/4)  & cc-pVQZ    & -7.98661  \ (88/32) \\
              & VB3        & -7.98669  \ (26/6)  & cc-pV5Z    & -7.98671  \ (134/42) \\
\hline
$10^{-4}$     & VB1        & -7.98658  \ (14/6)  & cc-pVTZ    & -7.98661  \ (58/29) \\
              & VB2        & -7.98693  \ (18/7)  & cc-pVQZ    & -7.98698  \ (88/44)  \\
              & VB3        & -7.98693  \ (26/9)  & cc-pV5Z    & -7.98696  \ (134/63) \\
\hline
$10^{-5}$     & VB1        & -7.98681  \ (14/8)  & cc-pVTZ    & -7.98678  \ (58/36) \\
              & VB2        & -7.98699  \ (18/9)  & cc-pVQZ    & -7.98701  \ (88/52) \\
              & VB3        & -7.98689  \ (26/13) & cc-pV5Z    & -7.98700  \ (134/74) \\
\hline
\hline
              & STO--basis & $E_{HF}^{STO}$ & GTO--basis    & $E_{HF}^{GTO}$ \\
\hline
              & VB1        & -7.98479       & cc-pVTZ      & -7.98695  \\
HF            & VB2        & -7.98650       & cc-pVQZ      & -7.98722  \\
              & VB3        & -7.98726       & cc-pV5Z      & -7.98733  \\
\hline
\hline
\end{tabular}

\end{table}

\newpage


\begin{table}
\centering

\caption{Total energies (Hartree) for the CO molecule using 
the HF and the xOEP methods using the STO and the GTO basis sets.
In the first column we indicate the value $\varepsilon$ 
used as a threshold in the TSVD decomposition of the 
TSVD decomposition of the $\mathbf{M M}^{\dagger}$ 
matrix in the xOEP method.
For the xOEP-STO and the xOEP-GTO results we have included, within 
brackets, the total number of occupied-virtual products 
(first number within the brackets) and the number of them that
are used (second number within the brackets) in each calculation. }
\label{TotalCO}

\begin{tabular}{|c|c|c|c|c|}
\hline
\hline
$\varepsilon$ & STO--basis & $E_{xOEP}^{STO}$ & GTO--basis & $E_{xOEP}^{GTO}$ \\
\hline
$10^{-3}$     & VB1        & -112.77681\ (93/10) & cc-pVTZ    & -112.77573\ (438/94) \\
              & VB2        & -112.78251\ (111/13)& cc-pVQZ    & -112.78246\
(686/104) \\
              & VB3        & -112.78433\ (184/21)& cc-pV5Z    & -112.78428\
(1046/116)\\
\hline
$10^{-4}$     & VB1        & -112.77999\ (93/16) & cc-pVTZ    & -112.77657\
(438/128) \\
              & VB2        & -112.78469\ (111/20)& cc-pVQZ    & -112.78392\
(686/176)  \\
              & VB3        & -112.78521\ (184/25)& cc-pV5Z    & -112.78519\
(1046/238) \\
\hline
$10^{-5}$     & VB1        & -112.77780\ (93/24) & cc-pVTZ    & -112.77704\
(438/160) \\
              & VB2        & -112.78524\ (111/27)& cc-pVQZ    & -112.78464\
(686/226) \\
              & VB3        & -112.78526\ (184/32)& cc-pV5Z    & -112.78551\
(1046/312) \\
\hline
\hline
              & STO--basis & $E_{HF}^{STO}$ & GTO--basis    & $E_{HF}^{GTO}$ \\
\hline
              & VB1        & -112.78199     & cc-pVTZ      & -112.78014  \\
HF            & VB2        & -112.78959     & cc-pVQZ      & -112.78891  \\
              & VB3        & -112.79056     & cc-pV5Z      & -112.79064  \\
\hline
\hline
\end{tabular}

\end{table}

\newpage


\begin{table}
\centering

\caption{Total xOEP energies (Hartree) for several atomic and 
diatomic systems calculated using a TSVD decomposition for 
the TSVD decomposition of the $\mathbf{M M}^{\dagger}$ 
matrix in the xOEP method (the threshold $\varepsilon$ used 
to discriminate the eigenvalues was $10^{-5}$). }
\label{TotalxOEP}

\begin{tabular}{|c|c|c|cc|}
\hline
\hline
 &    Numeric  & STO(VB3)   & GTO(cc-pV5Z)    &  GTO(Hess)\footnotemark[1] \\
\hline
Be  & -14.57254\footnotemark[2]   & -14.57254  &  -14.57255  &  -14.57243 \\
Ne  &-128.54553\footnotemark[3]   &-128.54540  & -128.54548  & -128.54538 \\
LiH  &  -7.98691\footnotemark[3]    &  -7.98689  &  -7.98700   &          \\
Li$_{2}$ & -14.87076\footnotemark[3]    & -14.87044  & -14.87090   &     \\
BH     & -25.12963\footnotemark[3]    & -25.12941  & -25.13013   &     \\
CO     &-112.785(3)\footnotemark[3]   &-112.78526  &-112.78551   & -112.78491 \\
\hline
\hline
\end{tabular}

\footnotetext[1]{Ref.~\cite{2007HGDSG}} 

\footnotetext[2]{Ref.~\cite{1993EV}}

\footnotetext[3]{Ref.~\cite{2009MKK}}

\end{table}

\newpage


\begin{table}
\centering

\caption{Orbital energies (a. u.) for the Be atom and the 
LiH and Li$_{2}$ molecules. 
Numerical results for the xOEP are presented. 
All STO (GTO) calculations have been done with the VB3 (cc-pV5Z) 
basis set.
The threshold $\varepsilon$ used in the TSVD to discriminate 
the eigenvalues of the $\mathbf{M M}^{\dagger}$ matrix in the 
xOEP method was $10^{-5}$. }
\label{OrbitalxOEP}

\begin{tabular}{|c|cc|ccc|}
\hline
\hline
  & HF-STO   & HF-GTO & xOEP-numeric   & xOEP-STO   &  xOEP-GTO \\
\hline
\hline
Be             &              &             &          &    &  \\      
\hline
1s        &  -4.17045    &  -4.17326   & -4.1668        &-4.1689  & -4.1704        \\        
2s        &  -0.34903    &  -0.34925   &-0.34885       &-0.34878  & -0.34892       \\
\hline
\hline
LiH    &       &       &         &          &      \\
\hline
$1 \sigma$   &  -2.44543    &  -2.44534   &-2.0786    &-2.09092  & -2.07071       \\
$2 \sigma$   &  -0.30172    &  -0.30172   &-0.3011    &-0.31391  & -0.31384       \\
\hline
\hline
Li$_{2} $      &         &        &          &           &       \\                     
\hline
$1 \sigma_{g}$  &  -2.45311    &  -2.44994   &-2.0276    &-2.00774  &-2.01361       \\
$1 \sigma_{u}$  &  -2.45311    &  -2.44994   &-2.0272    &-2.01262  &-2.00738       \\
$2 \sigma_{g}$  &  -0.18194    &  -0.18193   &-0.1813    &-0.18401  &-0.18616       \\
\hline
\hline
\end{tabular}

\end{table}

\newpage


\begin{table}
\centering

\caption{Decomposition of the total energy (Hartree) in its terms 
for the Be atom and the CO molecule.
All STO (GTO) calculations have been done with the VB3 (cc-pV5Z) 
basis set. The threshold $\varepsilon$ used in the TSVD to 
discriminate the eigenvalues of the $\mathbf{M M}^{\dagger}$ 
matrix in the xOEP method was $10^{-5}$. }
\label{Edecomp}

\begin{tabular}{|c|c|cc|cc|}
\hline
\hline
  &HF-Numeric & HF-STO       &   xOEP-STO     & HF-GTO     & xOEP-GTO \\
\hline
\hline
Be    &         &         &          &         &        \\
\hline
$T_{s}$    &     & 14.57301     &   14.57319     & 14.57301   & 14.57309  \\
$E_{nuc}$  &     &-33.63509     &  -33.63271     &-33.63518   & -33.63387  \\
$E_{bi}$ %
\footnote{This term is the sum of the Hartree and the exchange energies.}  
&   &  4.48911     &    4.48964     &  4.48916   &  4.48944   \\
$E_{tot}$ & -14.57304\footnote{Ref.~\cite{1992BBBC}}     &-14.57300     &  -14.57254   
 &-14.57301   & -14.57255    \\
\hline
\hline
CO     &           &          &           &        &       \\
\hline
$T_{s}$   &   & 112.64045    &  112.63314     & 112.64200  & 112.63348    \\
$E_{nuc}$ &   &-310.87673    & -310.86921     &-310.88020  &-310.86944    \\
$E_{bi}$ $^{\textit{a}}$ %
          &   & 62.93165     &  62.93421       & 62.93349  &  62.93637    \\
$E_{tot}$ & -112.79078\footnote{Ref.~\cite{2005Jensen}}   &-112.79056    &  -112.78526   
&-112.79064  &-112.78551    \\
\hline
\hline
\end{tabular}

\end{table}

\newpage 



\end{document}